\documentclass[12pt]{iopart}
\usepackage{amsmath}
\usepackage{physics}
\usepackage{graphicx}
\usepackage{hyperref}
\usepackage{cite}
\usepackage{orcidlink}
\usepackage{float}

\begin{document}

\title[Constraining parametric deviation from the Kerr spacetime]{Constraining parametric deviations from the Kerr spacetime using black hole ringdowns of GW150914 and GW190521}

\author{Zaryab Ahmed \orcidlink{0009-0009-6244-7565}$^{1,2}$, Shilpa Kastha \orcidlink{0000-0003-0966-1748}$^2$, Alex B. Nielsen \orcidlink{0000-0001-8694-4026}$^1$}

\address{$^1$ Department of Mathematics and Physics, University of Stavanger, NO-4036 Stavanger, Norway}
\address{$^2$ Niels Bohr International Academy, Niels Bohr Institute, Blegdamsvej 17, 2100 Copenhagen, Denmark}
 
\vspace{10pt}

\begin{abstract}

The ringdown phase of a binary black hole merger is modelled by the quasi-normal modes of a perturbed Kerr black hole. According to the black hole no-hair theorem, the emitted ringdown spectra are constrained by the mass and spin of the remnant black hole and thus offer an excellent test of the Kerr-nature of black holes. As a parameterization of beyond-Kerr effects, we employ the Johannsen-Psaltis metric ansatz. We analyse the ringdown of two binary black hole merger events-- GW150914 and GW190521 to constrain the deviation from Kerr. We find that both events are consistent with the Kerr metric and due to the larger signal-to-noise ratio and the presence of the additional subdominant mode in the ringdown phase, we find a factor of $\approx 2$ improvement in constraints on the deviation parameter in the case of GW190521 as compared to GW150914. Moreover, we find that the deviation parameter is anti-correlated to the spin of the remnant black hole. We also explore the effect of priors for other physical parameters on the Kerr deviation. 

\end{abstract}



\maketitle


\section{\label{sec:level1}Introduction\protect}
 
Compact binary dynamics is conventionally divided into three phases - the adiabatic inspiral with two distinct bodies, the rapid merger of these two bodies into one single body and the prompt ringdown of the final body to a stationary, quiescent state. In the case of black holes (BHs), the sweep through the inspiral and merger phases, as well as settling down to the final black hole is governed by the emission of gravitational waves (GWs).
    
At sufficiently later times beyond the merger, the emitted waveform can simply be written as a superposition of damped sinusoids, termed quasi-normal modes (QNMs). The entire spectrum consists of different fundamental modes and overtones defined by complex-valued frequencies labeled with three integers $\ell, m, n$ with $\ell\geq 2$, $-\ell\leq m \leq\ell$, and $n\geq 0$. Within general relativity (GR), the ringdown of the remnant object is modeled by a perturbed Kerr~\cite{Kerr:1963ud} black hole. Here, the QNM frequencies and their decay rates are computed from the linear analysis around the background Kerr geometry~\cite{Teukolsky:1971wp}. The Teukolsky master equation has to be solved to extract the QNMs. A large diversity of methods have been adopted to solve such master equations \cite{London:2014cma}.

Finding the QNM spectrum in a non-Kerr scenario is a non-trivial task and has been computed for only a small number of non-GR theories admitting stationary, spherically symmetric, vacuum solutions. 
It is easier to handle a single metric rather than a fully dynamic theory and to this end, Glampedakis et al.~\cite{Glampedakis:2017dvb} establish grounds for studying the ringdown in non-Kerr spacetimes. One can compute the photon orbit properties to generate the complex QNM templates 
\begin{align}
\label{eq:FrequencyDeviation}
    \omega = \omega_K + \delta\omega
\end{align}
where $\omega_K$ and $\omega$ are the QNM spectra of Kerr and non-Kerr spacetimes respectively and $\delta\omega$ encodes the deviations from the Kerr QNMs. A later study~\cite{Glampedakis:2019dqh}, shows that this technique can efficiently computes the QNMs of a non-Kerr spacetime.

In this technique, one can approximate the frequencies ($\mathrm{Re}(\omega)$) and damping times (${1}/{\mathrm{Im}(\omega)}$) of the GW ringdown signal by calculating the orbital frequency ($\Omega$) and Lyapunov exponent ($\gamma$) of photon (or graviton) orbits around a perturbed black hole. This is conventionally written as a complex frequency
\begin{align}
    \omega = l\Omega -\gamma(n+\frac{1}{2})~\iota.
\end{align}
The correspondence between unstable photon orbits and perturbations of the background spacetime, in the short-wavelength limit was investigated by Press~\cite{Press:1971wr} and later by Mashhoon and Ferrari~\cite{Ferrari:1984zz, Ferrari:1984ozr} and has since been studied in depth for Kerr black holes~\cite{PhysRevD.82.104003, PhysRevD.86.104006, PhysRevD.87.041502, PhysRevD.88.044047, Berti_2006, Cardoso:2008bp}.

In the case of a perturbed Kerr black hole, according to the no-hair theorem, mass and spin are the first two multipole moments of the spacetime, and all higher-order moments can be expressed in terms of these two moments. Consequently, the emitted ringdown spectra in such a scenario are completely and uniquely characterized by the final mass and the spin of the black hole. However, theories beyond GR, allow for black holes to differ from Kerr and may have additional ``hair". Probing such beyond GR theories demands a framework that considers spacetimes deviating from the Kerr metric.

To study such deviations, we make use of the Johannsen-Psaltis (JP) metric~\cite{Johannsen:2011dh} as an alternative description to the standard Kerr hypothesis. The JP metric is stationary and axisymmetric but not a vacuum solution to GR. Despite being a non-vacuum solution, it still parameterizes deviations from Kerr and has been used in the literature to constrain  these deviations~\cite{Bambi:2011ek,Bambi:2012pa,Kong:2014wha,Bambi:2015ldr,Glampedakis:2017dvb,Carson:2020iik,Dey:2022pmv}. The accuracy of the approximation we use here to produce the QNMs has been explicitly checked against numerical results for the JP metric~\cite{Glampedakis:2019dqh}. Using eikonal QNMs of non-GR metrics in tests of GR have been discussed in subsequent works including~\cite{Carullo:2021dui,Chen:2022nlw,Franchini:2023eda,LISAConsortiumWaveformWorkingGroup:2023arg}.

For a single ringdown mode, the two QNM parameters, frequency and damping time, can be used to constrain the mass and spin of a Kerr black hole. To constrain additional parameters, such as the deviation from Kerr in equation \ref{eq:FrequencyDeviation} requires additional information. One way of performing such a test, using black hole ringdowns alone, requires the detection of at least two ringdown modes~\cite{Dreyer:2003bv}. Recently, a number of authors have reported evidence of additional modes in the ringdown signal of black hole mergers~\cite{Isi:2019aib,Giesler:2019uxc,Capano:2021etf,LIGOScientific:2020tif,LIGOScientific:2021sio,Capano:2022zqm,Siegel:2023lxl, Wang:2023xsy,Gennari:2023gmx}. References~\cite{LIGOScientific:2020tif,LIGOScientific:2021sio} perform searches for overtones and higher modes based on methods developed in ref.~\cite{Carullo:2019flw}. References~\cite{Isi:2019aib, Giesler:2019uxc, Wang:2023xsy} discuss the evidence for the first overtone corresponding to the dominant mode of GW150914. However, the existence of the overtone so close to the merger time raises some interesting remarks regarding the data analysis techniques and theoretical issues~\cite{Cotesta:2022pci,Isi:2022mhy} of such a conclusion. 

Based on astrophysical assumptions about the total mass and mass ratio distributions of binary black hole systems in the observable universe, the general expectation was to observe only the dominant mode in the detections with the current generation of gravitational wave detectors~\cite{Berti:2016lat, Cabero:2019zyt}. Most of the detections till date~\cite{GW150914, GWTC1,GWTC2,KAGRA:2020tym, GWTC3, Nitz:2021zwj,Nitz:2021uxj, Mehta:2023zlk} by the Advanced LIGO\cite{Aasi_2015} and advanced Virgo\cite{Acernese_2015} detectors are found to be consistent with such expectations barring a few alternative scenarios. In the case of \cite{Capano:2021etf}, statistical evidence for an additional fundamental mode beyond the dominant mode was found in the event GW190521 \cite{LIGOScientific:2020iuh}. The additional mode was found in \cite{Capano:2021etf} and \cite{Capano:2022zqm} to be consistent with the Kerr hypothesis at high confidence. This implies that specific models leading to deviations of the form of equation (\ref{eq:FrequencyDeviation}) must be constrained. One question we want to address here is what the ringdown of GW190521 constrains.

In the eikonal approximation, the real part of the 220 mode and 330 mode are related by
\begin{align}
    \mathrm{Re}(\omega_{330}) = \frac{3}{2} \mathrm{Re}(\omega_{220})
    \label{eq:real}
\end{align}
and the imaginary parts are related by 
\begin{align}
    \mathrm{Im}(\omega_{330}) = \mathrm{Im}(\omega_{220}).
    \label{eq:imaginary}
\end{align}
Since the factors of $l$, $m$ and $n$ do not affect the orbital frequency and Lyapunov exponent of the photon orbits, the above relations hold in the eikonal approximation irrespective of the theory one is interested in. This is certainly true for the Kerr background and by the post-Kerr QNM toolkit of ref.~\cite{Glampedakis:2017dvb}, remains true for any other Kerr-like background as well.

A recent study~\cite{Dey:2022pmv} reported a ringdown analysis of the gravitational wave event GW150914 based on the JP metric ansatz. They compute the deviations of the QNM frequencies from the Kerr values, in the eikonal limit for the dominant $\ell=m=2,~n=0$ mode. They also perform a Bayesian analysis to find the posterior distribution on the quadrupolar deviation parameter. They find that the mass posterior is poorly constrained due to the presence of the deviation parameter, $\epsilon_3$. However, using further information from the inspiral part of the signal they find constraints on the same to be comparable to those from X-ray observations. They also perform a similar analysis for future detections of GW150914-like systems in the context of the Einstein Telescope. They used simulated GW150914-like systems including the 330 mode in the analysis, and show that constraints on deviations from the Kerr quadrupole could be obtained at percent level.

In this paper, we provide a comparative study of the constraint on JP metric parameters using the ringdown observation from the two GW events GW150914~\cite{LIGOScientific:2016vlm} and GW190521~\cite{LIGOScientific:2020iuh}. We compare our results for GW150914 with the previously reported bounds in ref.~\cite{Dey:2022pmv} and find broad agreement with their results. Moreover, we show the effects of prior choices in that case. Similarly, we carry out the same analysis for GW190521. Due to the large signal-to-noise ratio and the presence of the additional subdominant mode, we find a tighter constraint on the deviation parameter defined in the JP metric as compared to GW150914. Since we do not deal with a fully dynamic theory, the results of this work do not constitute a test of GR. Rather, we  concern ourselves with possible deviations at the metric level only.

This paper is organized as follows. 
In sec.~\ref{sec:JP} we discuss the JP space-time metric and present its photon orbit locations leading to quasi normal mode frequencies. In sec.~\ref{ringdown-analysis} we describe the methods used to obtain all the results reported in the following sec.~\ref{result}. Finally we provide concluding remarks in sec.~\ref{conclusion}.

\section{\label{sec:JP}Johanssen-Psaltis metric and circular photon orbits}

The JP metric is an infinite parameter family of stationary and axisymmetric metrics designed for tests of the no-hair theorem~\cite{Johannsen:2011dh}. It is conventionally written as
\begin{align}
    ds^2 &= -[1 + h(r,\theta)](1-\frac{2Mr}{\Sigma}) dt^2 -[1 + h(r,\theta)]\frac{4aMr \sin^2{\theta}}{\Sigma}dt d\phi \nonumber \\
    &+ \frac{\Sigma[1 + h(r,\theta)]}{\Delta + a^2\sin^2{\theta}h(r,\theta)} dr^2 + \Sigma d\theta^2 + \Big[\sin^2{\theta}\Big(r^2 + a^2 +\frac{2a^2 Mr \sin^2{\theta}}{\Sigma}\Big) \nonumber \\
    &+ h(r,\theta) \frac{a^2(\Sigma + 2Mr)\sin^4{\theta}}{\Sigma}\Big] d\phi^2
\end{align}
where $\Sigma \equiv r^2 + a^2~\cos^2{\theta} $ and $\Delta \equiv r^2 - 2Mr +a^2$ are the usual functions from Kerr. The function
\begin{align}
    h(r,\theta) = \sum^{\infty}_{k=0}\Big(\epsilon_{2k} + \epsilon_{2k+1} \frac{Mr}{\Sigma}\Big)\Big(\frac{M^2}{\Sigma}\Big)^k
\end{align}
accounts for parametric deviations from the Kerr metric~\cite{Johannsen:2011dh} and contains an, in principle, infinite set of dimensionless deviation parameters from Kerr, denoted by $\epsilon_k$. The Kerr metric in Boyer-Lindquist coordinates is recovered in the case of $h(r,\theta)=0$. If the coordinate $t$ matches the proper time of observers at infinity, the requirements of asymptotic flatness demand $\epsilon_0 = \epsilon_1 = 0$. $\epsilon_2$ is constrained by the Lunar Ranging experiments~\cite{Williams:2004qba} to  $|\epsilon_2|<4.6 \times 10^{-4}$. The next non-zero, unconstrained parameter is $\epsilon_3$ and in this work, for simplicity, we restrict $h(r,\theta)$ to be just
\begin{align}
    h(r,\theta) = \epsilon_3 \frac{M^3 r}{\Sigma^2}
\end{align} which provides parameterized perturbations to the Kerr metric depending on a single parameter $\epsilon_3$. Geometrically, the outer horizon is more prolate for positive values of $\epsilon_3$ and more oblate for negative values, with respect to the outer horizon for a Kerr black hole. This choice of $h(r,\theta)$ also corresponds to quadrupolar deformations of the form~\cite{Glampedakis:2017dvb}
\begin{align}
    Q_{JP} = Q_K + \epsilon_3M^3
\end{align}
where $Q_K$ is the Kerr quadrupole. As highlighted in ref.~\cite{Glampedakis:2017dvb}, the JP metric with $\epsilon_3$ is not a vacuum solution of the Einstein equations and care must be taken in interpreting multipole decompositions. Nonetheless, for stationary and axisymmetric spacetimes, the JP metric provides a convenient parameterization of non-Kerr effects in the black hole ringdown phase.
\begin{figure}
    ~\centering
    \includegraphics[width=0.6\textwidth]{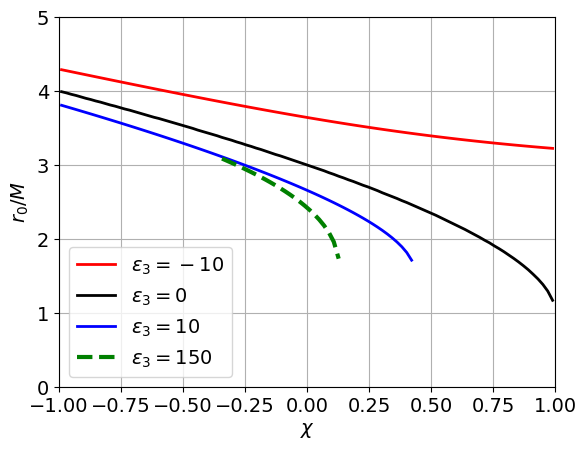}
   ~\caption{Dimensionless coordinate locations of the equatorial photon orbits in the JP geometry for $\epsilon_3 = -10, 0, 10$  and a high value, $\epsilon_3 = 150$. The dashed, green line denotes $\epsilon_3 = 150$ and shows that photon orbits can exist for $\epsilon_3>80$, which is the upper bound on the $\epsilon_3$ prior in ref.~\cite{Dey:2022pmv}.}
    \label{fig:LRpos}
\end{figure}
The radial locations of equatorial photon orbits can be calculated by finding the roots of the effective potential $V_{eff}$ and its derivative $V'_{eff}$ (Eqns. 81 and 80, respectively, of~\cite{Glampedakis:2017dvb}). This system of equations does not admit an analytic solution and needs to be solved numerically. Fig. 1 of~\cite{Glampedakis:2017dvb} shows the dimensionless radial locations of photon orbits for $|\epsilon_3| = 0.1,~ 1,~ 10$ against the dimensionless spin parameter $\chi=a/M$. 

The shaded region of Fig. 2 from~\cite{Johannsen:2011dh} shows the region of the parameter space in $(\epsilon_3,~\chi)$ where the central object is a naked singularity. Outside the shaded region, the central object is a black hole and one can expect light rings to exists for any combination of $(\epsilon_3,~\chi)$ outside the shaded region. The same figure shows that for higher values of $\epsilon_3$, the spins cannot be too large in order for an event horizon (and photon orbit) to exist. This feature is also observed in our Figure.~\ref{fig:LRpos} where we include the photon orbit locations for $\epsilon_3 = 150$, given the dimensionless spins, $\chi =[-1,1]$. The dashed, green line in Figure.~\ref{fig:LRpos} does not extend to $\chi=-1$ or $\chi=1$ because the high value $\epsilon_3 = 150$ requires $\chi$ values to be near to $\chi=0$ in order to remain outside the shaded region in Fig. 2 of~\cite{Johannsen:2011dh}. 

Once the photon orbit location is found, the dimensionless orbital frequencies and Lyapunov exponent can be derived using the metric. This can be achieved for each relevant location in the JP parameter space, without making any leading order approximation, keeping the full solutions. (A linearized version of the same analysis can be found in references~\cite{Glampedakis:2017dvb,Carson_2020,MasterThesis:2023}.) The orbital frequency and Lyapunov exponent can then be converted to ringdown frequencies and damping times, which themselves can be compared to the observed data.

\begin{figure}
    ~\centering
    \includegraphics[width=0.7\textwidth]{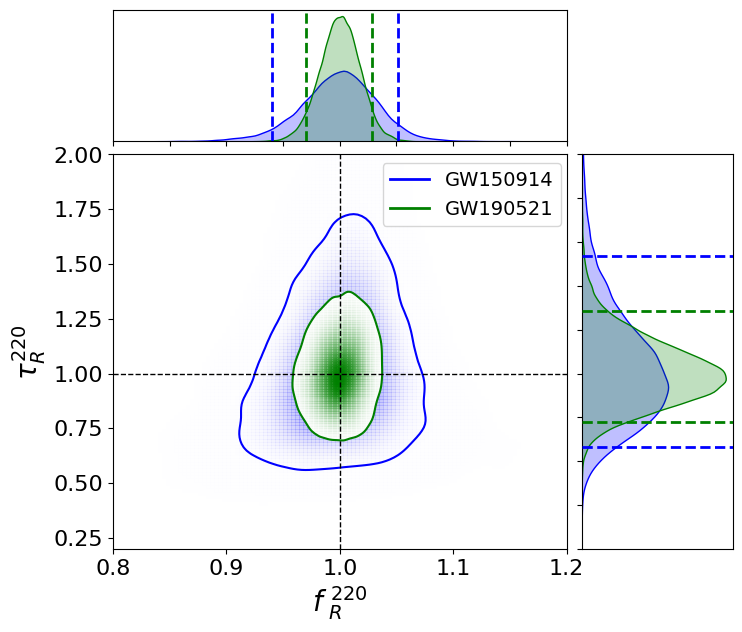}
   \caption{Posterior distribution for dimensionless ringdown frequencies and damping times of the two events GW150914 and GW190521. $f^{220}_R$ and ${\tau}^{220}_R$ refer to the $f$ and $\tau$ distributions divided by their respective median values. The blue and green colors refer to the 90\% confidence contours of the $(f^{220}_R,\tau^{220}_R)$ posterior distributions for the events GW150914 and GW190521, respectively.}
    \label{fig:SNR}
\end{figure}

\section{Ringdown Analysis}\label{ringdown-analysis}

\begin{figure}
    ~\centering
     \includegraphics[scale=0.6]{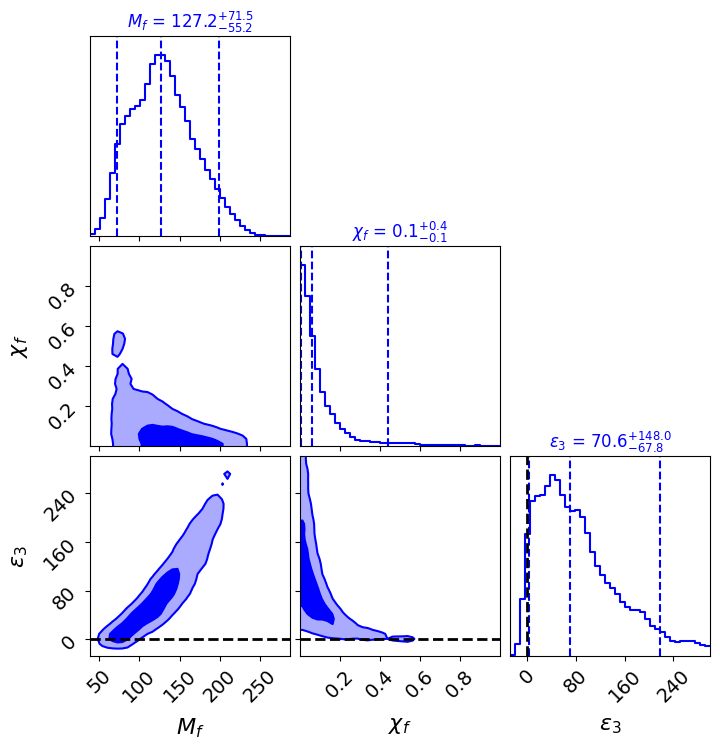}
    ~\caption{Posterior distributions obtained from GW150914 ringdown data using uniform priors; $M_f=[20,300] M_{\odot}$, $\chi_f=[0,0.99]$ and $\epsilon_3=[-30,300]$. The Kerr deviation parameter $\epsilon_3$ is positively correlated with the mass $M_f$ and negatively correlated with the spin $\chi_f$. The vertical dashed lines correspond to the 5\%, median and 95\% values of the parameters. The dark and light blue regions correspond to 50\% and 90\% confidence regions, respectively.}
     \label{fig:Full}
 \end{figure}

The GW ringdown signal is studied by fitting damped sinusoids to the strain data. Each damped sinusoid is parameterized by a single frequency and a single damping time. Such an agnostic comparison of damped sinusoids to the data has been performed for GW190521 in in~\cite{Capano:2021etf} using the PyCBC package~\cite{Biwer:2018osg}. Here we additionally perform the same analysis for GW150914 using the same method as~\cite{Capano:2021etf} for consistency (available at~\cite{GIT}).  

An advantage of performing a ringdown-only analysis on GW190521 instead of GW150914, even on just a single 220 mode, is the greater signal-to-noise ratio $\sim$ 12.2~\cite{Capano:2021etf} in the ringdown phase for GW190521 compared to GW150914 ringdown signal-to-noise ratio $\sim$ 8.5~\cite{LIGOScientific:2016lio}. The full IMR signal-to-noise ratio of GW150914 is greater than that of GW190521, but because of its greater apparent mass, very little inspiral signal is seen for GW190521 and its ringdown phase is closer to the peak sensitivity of the detectors. This means that the relative frequency and damping time of the 220 ringdown mode is more tightly constrained for GW190521, as depicted in Figure.~\ref{fig:SNR}. A possible disadvantage of using GW190521 to constrain $\epsilon_3$ is that little inspiral signal is visible in the data, although this is not relevant to our analysis here, which is restricted to the ringdown phase only.

A key issue in ringdown analyses is the choice of start time, since signal from the merger can bias ringdown-only inference. A reasonable choice of starting time is generally considered to be $\sim 10-15M$ after merger~\cite{Kamaretsos:2012bs,Bhagwat:2017tkm,Cabero:2017avf,Carullo:2018sfu,Carullo:2019flw}, considerably away from the merger time to avoid contamination of the ringdown-only signal. For GW150914, we start our ringdown analysis $3\mathrm{ms}$ after merger $t = t_M + 3\mathrm{ms}$ where $t_M = 1126259462.423 s$ is the GPS time at the Hanford detector~\cite{LIGOScientific:2016lio}.  In contrast, the starting time for the ringdown analysis of~\cite{Dey:2022pmv} is directly at merger. The analysis of~\cite{LIGOScientific:2016lio} suggests that for GW150914, $3\mathrm{ms}$ after merger is a reasonable time to study a single ringdown mode, since at this time the $3\mathrm{ms}$ damped sinusoid parameters are consistent with those expected from a full Inspiral-Merger-Ringdown (IMR) analysis in the GR model. For GW190521, we choose $6\mathrm{ms}$ after merger with $t_M = 1242442967.445 s$ to match the choice of~\cite{Capano:2021etf}.

The computation of QNMs of the Kerr background itself has some degree error due to the use of approximations in solving the highly non-linear system. For example, ref.~\cite{Yang:2012he} report the errors for the fundamental modes when using the WKB method. QNM data provided in reference~\cite{Berti:2005ys} takes care of such errors by using numerical fits. We employ the offset function (and the corresponding fitting coefficients) of ref.~\cite{Glampedakis:2017dvb} to account for such errors when producing the frequencies and damping times of the JP background. 

Following ref.~\cite{Glampedakis:2017dvb} we employ an offset function and the corresponding fitting coefficients for non-Kerr QNMs to ensure that the errors of the Kerr QNMs are maintained. Therefore, we use the same fitting coefficients when producing the frequencies and damping times of the JP background. 

In the JP model, the frequencies $(f^{JP})$ and damping times $(\tau^{JP})$ of the damped-sinusoid QNMs are functions of the three parameters, $(M_f,~\chi_f,~\epsilon_3)$ obtained using equations 36 and 47 of ref.~\cite{Glampedakis:2017dvb}. Each allowed combination of $(M_f,~\chi_f,~\epsilon_3)$ produces a pair $(f^{JP},~\tau^{JP})$. The relations between $(f^{JP},~\tau^{JP})$ and $(M_f,~\chi_f,~\epsilon_3)$ are in general non-linear for large $\epsilon_3>1$ and we employ a root-finding algorithm to invert these relations.

We adopt a probability distribution for $(f,\tau)$ using the freely available agnostic data samples of~\cite{Capano:2021etf} and also for GW150914 agnostic data samples prepared in a similar fashion. We then evaluate the probability distribution at each of the computed JP pairs $(f^{JP},~\tau^{JP})$ to find its likelihood. JP pairs within the probability distribution return a non-zero value, $k$, which is then used to generate a density of the parameter by storing the respective combination  $k$-times, which we ultimately call the posterior density. For a sufficient number of samples drawn from a uniform prior distribution of each of the parameters $(M_f,~\chi_f,~\epsilon_3)$, we can then generate a posterior distribution of each of these parameters based on the likelihoods of the computed $(f,\tau)$ pairs.

We do not include the amplitude $A_{lmn}$ or the phase $\phi_{lmn}$ of QNMs in our analysis, since they do not have an impact on the frequencies and damping times, irrespective of the background metric. This independence is also shown in ref.~\cite{Dey:2022pmv}. This means that the bounds on $(M_f,~\chi_f,~\epsilon_3)$ depend only on the QNM parameters $(f^{JP},~\tau^{JP})$.

Since for a single QNM, this procedure obtains distributions for three parameters, $(M_f,~\chi_f,~\epsilon_3)$, from just two input parameters $(f,~\tau)$, the resultant parameters are highly correlated. In particular, this means that priors and prior boundaries on the parameters $(M_f,~\chi_f)$ affect the constraints obtained for $\epsilon_3$. To study the effect of this, we show in Figs.~\ref{fig:Full} and~\ref{fig:restricted} the effect for GW150914 of two different potential ways of choosing the mass prior.

In Figure.~\ref{fig:Full} we have fixed flat priors in $\chi_f=[0,0.99]$ and $\epsilon_3=[-30,300]$ and extended the mass prior to $M_f=[20,300]M_{\odot}$ such that the marginalized $\epsilon_3$ posterior is well contained within its prior range, without railing against the range boundaries. This choice would correspond to the most agnostic, ringdown-only choice.
 
Another possible choice is to choose mass priors based on other astrophysical grounds. This is similar to the approach adopted in~\cite{Dey:2022pmv}. As an example of this, in~\ref{fig:restricted} we used the maximum likelihood of the detector frame IMR mass of GW150914 and GW190521 (from references~\cite{LIGOScientific:2016vlm} and~\cite{LIGOScientific:2020ufj} respectively), such that the prior on the final mass is $M_f = [0.5,1.5]M^{IMR}$. We discuss the effects of different prior ranges in section~\ref{result}. One could in principle construct other prior distributions for the masses, based on the distribution of black hole masses observed by other means. But the choice of our example here gives us a simple way to compare the two gravitational wave events, GW150914 and GW190521.

\section{Results}
\label{result}

Performing the analysis described in section~\ref{ringdown-analysis}, we
find that the parameters $(M_f,~\chi_f,~\epsilon_3)$ are strongly correlated and the prior choices significantly affect the obtained results. Figure.~\ref{fig:Full} is obtained for a single ringdown mode in GW150914 by allowing the $M_f$ and $\epsilon_3$ priors to be wide enough such that the posterior is well contained within the prior ranges.

We find that GW150914 data allows for $\epsilon_3$ up to $\approx 250$ corresponding to an upper bound on the $M_f \approx 200  M_{\odot}$. The data favors lower values of $\chi_f$ (higher values of $\epsilon_3$) and this too affects the measured values. We have restricted $\chi_f$ to be non-negative, but if $\chi_f < 0 $ are allowed, then $\epsilon_3$ is even less constrained and the posterior density supports even higher values of $\epsilon_3$. This anti-correlation between the spin value and the value of $\epsilon_3$ can be understood in terms of the behaviour of the photon orbit location in Figure.~\ref{fig:LRpos} and hence its frequency. The spin prior is kept fixed as $\chi_f = [0,0.99]$ throughout our analysis to maintain consistency with ref.~\cite{Dey:2022pmv}.

Figure.~\ref{fig:Full} is a broader version than that reported as Fig. 2 in ref.~\cite{Dey:2022pmv}. We have included here the full range of values of $\epsilon_3$ for which a photon orbit can exist and can be compatible with the data. Our figure represents the maximal constraint on $\epsilon_3$ that can be obtained from a ringdown-only analysis of a single mode in GW150914.

Although not shown here, results obtained using the same approach for GW190521 ringdown data show similar features for the three parameters $(M_f,~\chi_f,~\epsilon_3)$. Final spin values close to zero are preferred, the final mass takes values up to $1000 M_{\odot}$ and $\epsilon_3$ shows a wide range, similar to GW150914 when priors wide enough to exhaust support in the posterior are allowed.

It is also clear from Figure.~\ref{fig:Full} that the JP model favors a final spin $\chi_f$ of the JP remnant close to zero. This is due to the strong anti-correlation between $\chi_f$ and $\epsilon_3$. The effect of higher values of $\epsilon_3$ can be offset by bringing the spin down to very small values. Astrophysically we expect a remnant produced in a binary coalescence to have inherited significant angular momentum from the inspiral phase, both from the initial spins of the coalescing black holes and their orbital angular momentum. The median final mass value in Figure.~\ref{fig:Full} is also rather high. For example, it is much higher than that allowed by even a very rudimentary Peters and Mathews \cite{Peters:1963ux} type  analysis of the inspiral phase \cite{LIGOScientific:2016wyt}. 

\begin{figure}
    ~\centering
     \includegraphics[scale=0.6]{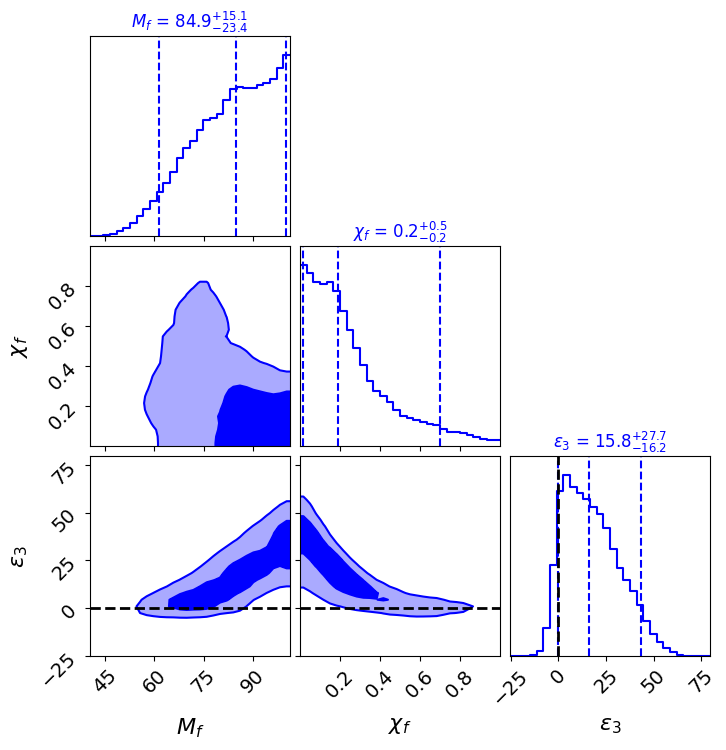}
    ~\caption{Same as Figure.~\ref{fig:Full}, showing posteriors on the parameters $M_f$, $\chi_f$ and Kerr deviation parameter $\epsilon_3$ but this time with IMR-informed mass prior: $M_f = [0.5,1.5]\times67M_{\odot}$.}
     \label{fig:restricted}
 \end{figure}

Imposing more astrophysically motivated priors on $M_f$ and $\chi_f$ is qualitatively similar to the approach adopted in~\cite{Dey:2022pmv}. For GW150914 and GW190521, the maximum likelihood values obtained for detector frame masses are $M^{IMR}\approx67 M_{\odot}$~\cite{LIGOScientific:2016vlm} and $M^{IMR}\approx266 M_{\odot}$~\cite{LIGOScientific:2020ufj}, respectively. Choosing these IMR-informed mass priors, $M_f = [0.5,1.5]M^{IMR}$, in this way also allows for a like-to-like comparison of the measured values between the two events GW150914 and GW190521. Figure.~\ref{fig:restricted} shows the results based on such a restricted mass prior for GW150914. Our results are broadly consistent with results reported in~\cite{Dey:2022pmv} with minor differences attributed to the choices of ringdown start time and specific mass priors. $\epsilon_3$ is constrained below $70$, the remnant still shows some moderate degree of spin and the final mass is more astrophysically reasonable. We report the median value $\epsilon_3 = 15.8^{+27.7}_{-16.2}$ where the 5\% and 95\% values span an interval containing the Kerr mass and spin values when $\epsilon_3=0$.

\begin{figure}
    ~\centering
     \includegraphics[scale=0.6]{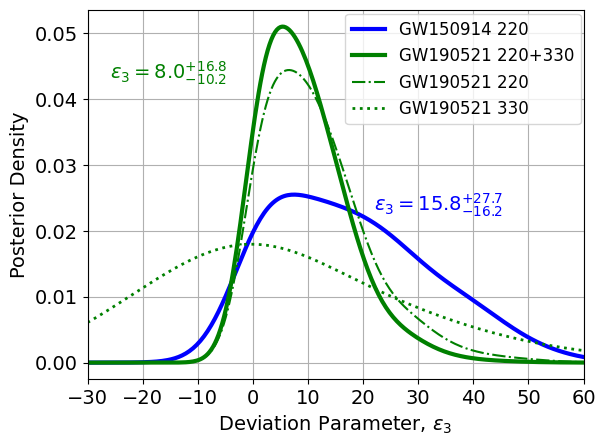}
    ~\caption{Comparative plot of posterior distributions for the deviation parameter, $\epsilon_3$. Solid lines show the posterior distributions using just the dominant mode of GW150914 (blue) and 220+330 ringdown modes of GW190521 (green). Median, 5\%, and 95\% values corresponding to these two posteriors are also noted by the text in their respective colours. The Kerr case, $\epsilon_3= 0$ is contained within the $90\%$ credible interval of both events. The  Dash-dotted and dotted green lines show the posteriors obtained using only the 220 and 330 mode of GW190521 respectively.}
     \label{fig:compare}
 \end{figure}

The same analysis using GW190521 ringdown data shows an improved constraint on the deviation parameter $\epsilon_3$ by roughly by a factor of two over GW150914, see Figure.~\ref{fig:compare}. This is due to the higher ringdown SNR of GW190521 compared to GW150914.

To this point, we have only applied our ringdown analysis to the single 220 mode. GW190521 however, shows evidence of a second fundamental mode, the 330 mode \cite{Capano:2021etf}.
Posteriors for $(M_f,~\chi_f,~\epsilon_3)$ can be obtained for the 220 mode and 330 mode separately and then combined together to obtain a composite distribution. The GW190521 220 mode on its own already gives an improvement relative to the GW150914 220 mode. Unfortunately, the ringdown parameters of the GW190521 330 mode are not well constrained, especially its damping time. The error bars on the $(f^{330},\tau^{330})$ measurement are reflected in the error bars of the resulting $\epsilon_3$ measurement, see Figure.~\ref{fig:compare}. However, because of the relations~(\ref{eq:real}) and~(\ref{eq:imaginary}) the composite posterior shows slight improvement compared to the posterior obtained using just 220 mode. This improvement is expected to be further enhanced if for a future event, the 330 mode is better constrained, or even additional modes are measured.

Comparative results for the two events are collected in Figure.~\ref{fig:compare}. The combined posterior density of $\epsilon_3$ using the 220 and 330 mode of GW190521 gives 90\% credible interval $\epsilon_3 = 8.0^{+16.8}_{-10.2}$, where again the 90\% range includes the Kerr mass and Kerr spin when $\epsilon_3 =0$. For the 220 mode only results of the two events, the lack of support in the posterior densities for $\epsilon_3<-10$ is due to the outer horizon of the JP remnant becoming more and more oblate, pushing the photon orbits further and further out (see Figure.~\ref{fig:LRpos}). The orbital frequencies of such photon orbits are too low to be consistent with the observed GW frequencies. The fall-off on the RHS of the peaks is dictated by the upper bound on the mass priors and the lower bound on the spin priors. This is also clear if one compares the $\chi_f\mathrm{~vs~}M_f$ plots of Figure.~\ref{fig:Full} and Figure.~\ref{fig:restricted}.

\section{Discussion and Conclusion}\label{conclusion}

We have presented constraints on the JP parameter $\epsilon_3$ that can be obtained from the ringdown phase of the two gravitational wave events GW150914 and GW190521. We have investigated the role that prior assumptions on the final mass parameter plays in constraining $\epsilon_3$. We have shown how tighter bounds can be obtained using the multimode detection GW190521, than have been obtained in the literature to date using just GW150914. We obtain $\epsilon_3=8.0^{+16.8}_{-10.2}$, using LVK ringdown data for GW190521, which is $\approx 2$ times better than a similar analysis on just GW150914. There are two factors that lead to this improvement; the higher signal-to-noise ratio of the ringdown for GW190521 and the existence of the subdominant mode. We expect that even tighter constraints could be obtained by combining these two events together. One could also use the relative amplitude of different modes to constrain the total mass of the system. However, such an analyses are beyond the scope of this work. 

We have also shown explicitly how an agnostic ringdown data file as described in section~\ref{ringdown-analysis} can be post-processed to provide constraints on specific models. While details will depend on the specific model considered, access to an existing agnostic data file is significantly useful. This avoids repeating computationally expensive Bayesian inference on GW data. Once such posterior sample results are available, it is easier and cheaper to perform further data analysis tasks without the need to re-fit the entire strain data.

Our analysis here has been restricted to the ringdown phase only. But this of course, can be further extended. Performing a full IMR test could give access to more information and therefore better estimates of the parameter values, if a significant part of the signal is observed outside the ringdown phase. Ref.~\cite{Carson:2020iik} uses LISA~\cite{Babak:2021mhe} and Cosmic Explorer~\cite{Reitze:2019iox} IMR synthetic data of GW150914-like events to show that the $\epsilon_3$ parameter can in fact be constrained to values of $\mathcal{O}(10^{-2})$. We note that our GW190521 ringdown-only result and the IMR combined estimate, quoted in ref~\cite{Carson:2020iik} as LIGO O2 GW150914, are of comparable size. This comparison should be seen in the context of the approximations used to obtained the QNMs of the JP metric. 

Ref.~\cite{Dey:2022pmv} also uses simulated GW150914 ringdown signals for the Einstein Telescope~\cite{Punturo:2010zz} (and also multimode injections) to show that the constraint on the JP (and the Manko-Novikov spacetime) deviation parameters is significantly improved for future detectors. The future prospects for improved constraints on the Kerr nature of black holes are bright indeed.

 
\ack
We thank Kallol Dey for sharing data and helpful discussions. S.K. and Z.A. thank Vitor Cardoso, and Gregorio Carullo for fruitful discussions. We also thank Vasco Gennari for a critical reading of the manuscript and for providing useful comments. S.K. and Z.A. acknowledge support from the Villum Investigator program supported by the VILLUM Foundation (grant no. VIL37766) and the DNRF Chair program (grant no. DNRF162) by the Danish National Research Foundation. This project has received funding from the European Union's Horizon 2020 research and innovation programme under the Marie Sklodowska-Curie grant agreement No 101131233. 

This research has made use of data or software obtained from the Gravitational Wave Open Science Center (gwosc.org), a service of the LIGO Scientific Collaboration, the Virgo Collaboration, and KAGRA. This material is based upon work supported by NSF's LIGO Laboratory which is a major facility fully funded by the National Science Foundation, as well as the Science and Technology Facilities Council (STFC) of the United Kingdom, the Max-Planck-Society (MPS), and the State of Niedersachsen/Germany for support of the construction of Advanced LIGO and construction and operation of the GEO600 detector. Additional support for Advanced LIGO was provided by the Australian Research Council. Virgo is funded, through the European Gravitational Observatory (EGO), by the French Centre National de Recherche Scientifique (CNRS), the Italian Istituto Nazionale di Fisica Nucleare (INFN) and the Dutch Nikhef, with contributions by institutions from Belgium, Germany, Greece, Hungary, Ireland, Japan, Monaco, Poland, Portugal, Spain. KAGRA is supported by Ministry of Education, Culture, Sports, Science and Technology (MEXT), Japan Society for the Promotion of Science (JSPS) in Japan; National Research Foundation (NRF) and Ministry of Science and ICT (MSIT) in Korea; Academia Sinica (AS) and National Science and Technology Council (NSTC) in Taiwan.


\section*{References}
\bibliographystyle{iopart-num}
\bibliography{iopart-num}

\end{document}